\documentclass[12pt,aps,prb,preprint]{revtex4}
\usepackage{graphicx}
\usepackage{cranm_timesv}

\usepackage{amsmath}

\begin{document}

\title{New views of the solar wind with the Lambert W function}

\author{Steven R. Cranmer}
\email{scranmer@cfa.harvard.edu}
\affiliation{Harvard-Smithsonian Center for Astrophysics,
Cambridge, Massachusetts 02138}

\begin{abstract}
This paper presents closed-form analytic solutions to two
illustrative problems in solar physics that have been
considered not solvable in this way previously. Both the
outflow speed and the mass loss rate of the solar wind of
plasma particles ejected by the Sun are derived analytically
for certain illustrative approximations. The calculated radial
dependence of the flow speed applies to both Parker's
isothermal solar wind equation and Bondi's equation of spherical
accretion. These problems involve the solution of transcendental
equations containing products of variables and their logarithms.
Such equations appear in many fields of physics and are solvable by
use of the Lambert $W$ function, which is briefly described. This
paper is an example of how new functions can be applied to
existing problems.

\hspace*{0.01in}

\noindent
{\em American  Journal of Physics,} in press $\,$(2004)
\end{abstract}

\maketitle

\section{Introduction}

\baselineskip=15pt
\parskip=0.113in
Most stars eject matter from their atmospheres and fill the
surrounding space with hot, low-density gas.\cite{LC99} Astronomers and space
physicists have studied the continuously expanding solar wind of charged
particles from the
Sun for almost a half century.\cite{Hu72,KR95,GP97} The study of the
solar wind as a unique plasma laboratory is compelling for several reasons.
Many basic processes in various fields of physics (for example, plasma
physics, electromagnetic wave theory, and nonequilibrium
thermodynamics) have been detected in the solar wind and almost
nowhere else. The solar wind is the closest example of a stellar wind,
and stellar winds affect the long-term evolution of galaxies by injecting
large amounts of matter and energy into the interstellar medium. On
the more practical side, when solar wind particles impact the
Earth's magnetosphere, they can interrupt communications, threaten
satellites and the safety of orbiting astronauts, and disrupt
ground-based power grids.\cite{SW01}

The crown-like solar corona seen during a total eclipse is
the place where the solar wind undergoes its initial
acceleration.
The shimmering auroras seen in northern and southern skies 
are the end products of the interaction between incoming
solar wind particles and the Earth's magnetic field.
Sightings of the corona and the aurora go back into antiquity,
but the first scientific understanding of the solar wind
came at the beginning of the 20th century.
Researchers gradually realized that there were strong
correlations between the appearances of sunspot activity,
geomagnetic storms, auroras, and motions in comet tails. In 1958,
Eugene Parker\cite{P58,P63} combined these empirical clues with the
knowledge that the bright solar corona consists of extremely hot
($10^6$\,K) plasma and postulated a model of a steady-state outward
expansion from the Sun. Parker's key insight was that the high
temperature of the coronal plasma provides enough energy per
particle to overcome gravity and produce a natural transition from a
subsonic (bound, negative total energy) state near the Sun to a
supersonic (outflowing, positive total energy) state in
interplanetary space. This theory was controversial at the time,
but Parker had only to wait four years until the existence of the
continuous, supersonic solar wind was verified by the Mariner
2 probe in 1962.\cite{Hu72}

Over the past decade, our understanding of the physics of
the solar wind has increased dramatically from both
new space-based observations and the rapid growth of
computer power for simulations
(see Refs.~\onlinecite{Jk97,As01,Cr02} for recent reviews). The Ulysses
spacecraft, for example, was the first probe to venture far from the ecliptic
plane and soar over the solar poles to measure the solar wind in three 
dimensions.\cite{Ma01} Remote observations of the solar corona have become
significantly more detailed with data from space-based telescopes pouring in
as never before. Figure~1 shows a snapshot of the corona as observed in 1996
by two instruments on the SOHO (Solar and Heliospheric Observatory)
spacecraft. However, progress in solar wind research often requires
substantial numerical analysis, because even the most basic problems have not
been tractable by analytic means. This paper takes advantage of a new
transcendental function that is unfamiliar to many physicists---the Lambert
$W$ function---to illustrate how two fundamental solar wind problems can be
solved analytically.

In this era of efficient numerical computation, it is worthwhile to
list the various ways that analytic solutions for the properties of
the solar wind can be useful (beyond their pure aesthetic appeal).
Analytic expressions often are used as initial guesses for more
complicated iterative, time-dependent, or multi-dimensional
calculations. Closed-form solutions also make it easier to study
linearized perturbations to a known background state. The rapid
evaluation of a large number of cases is facilitated by having
analytic formulae, especially because many symbolic computation
packages already contain optimized routines for the Lambert $W$
function. Finally, the ability to write down simple expressions for
solar wind plasma properties may make the extrapolation to
other stars more tractable and physically understandable.

This paper presents a brief overview of the Lambert $W$
function in Sec.~II and a summary of the governing
equations of the solar wind in Sec.~III.
The use of the $W$ function in solving the classical
Parker solar wind problem, that is, the radial dependence
of the wind speed for an isothermal plasma, is given
in Sec.~IV.
The use of this function in solving for the mass loss
rate of the solar wind is given in Sec.~V.
Conclusions and other potential applications of this
function are given in Sec.~VI.

\section{Some Properties of the Lambert W Function}

Like many mathematical functions, the Lambert $W$ function
was derived and used independently by several researchers
before the mathematics and computer science community settled
on a common notation in the mid-1990s.\cite{Co96}
This function has been used to solve problems in electrostatics, statistical
mechanics, general relativity, radiative transfer, quantum chromodynamics,
combinatorial number theory, fuel consumption, and
population growth,\cite{Co96,MO97,Ve00,Mg00,Ca03}
but is still not widely known by physicists.

The Lambert $W$ function is defined as the multivalued inverse
of the function $x e^{x}$.
Equivalently, the multiple branches of $W$ are the multiple
roots of the equation
\begin{equation}
W(z) e^{W(z)} = z ,
\end{equation}
where $z$ is in general complex. There are an infinite number of solution
branches, labeled by convention by an integer subscript: $W_{k} (z)$, for $k
= 0, \pm 1, \pm 2,
\ldots$ If $z$ is a real number $x$, the only two branches that take on real
values are $W_{0}(x)$ and $W_{-1}(x)$. Figure~2 plots these two branches,
which are the only ones that are needed in the applications of this paper.

Numerous formulae for the differentiation, integration,
and series expansion of $W$ are given in the references cited
above (for example, Ref.~\onlinecite{Co96,Ve00}).
One useful result, which is applied in the following, is given
here.
Near the branch cut point at $x = -1/e$, $W_{0} = W_{-1} = -1$, and
the two real branches can be approximated to lowest order by
\begin{eqnarray}
W_{0} (x) & \approx & -1 + \sqrt{2 + 2ex}
\label{eq:W0approx} \\
W_{-1} (x) & \approx & -1 - \sqrt{2 + 2ex}
\label{eq:W1approx}.
\end{eqnarray}

A useful way for expressing the solutions to a standard family
of transcendental
equations in terms of the Lambert $W$ function is to note that the
equation\cite{Br98}
\begin{equation}
\ln (A + Bx) + Cx = \ln D ,
\label{eq:briggs1}
\end{equation}
where $A$, $B$, $C$, and $D$ do not depend on $x$,
has the exact solution
\begin{equation}
x = \frac{1}{C} W \Big[ \frac{CD}{B}
\exp \big(\frac{AC}{B} \big) \Big] - \frac{A}{B}.
\label{eq:briggs2}
\end{equation}
The choice of solution branch usually depends on physical
arguments or boundary conditions.

\section{Governing Equations of the Solar Wind}

The expansion of the solar wind is treated traditionally as
a problem of steady-state hydrodynamics.
The Sun has a strong magnetic field that confines and
directs the flow of plasma (see Fig.~1), but along the
magnetic ``flux tubes,'' the dynamics is essentially independent of
the strength of the field. Thus, the overall properties of the
solar wind can be determined by solving the hydrodynamic equations
of mass, momentum, and energy conservation.\cite{Hu72,LL87}

The equation of mass conservation for a single-component fluid is given by
\begin{equation}
\frac{\partial \rho}{\partial t} + 
\nabla \cdot (\rho {\bf u}) = 0,
\label{eq:mascon}
\end{equation}
where $\rho$ is the mass density and ${\bf u}$ is the 
velocity.
The equation of momentum conservation is
\begin{equation}
\frac{\partial {\bf u}}{\partial t} +
({\bf u} \cdot \nabla) {\bf u} = 
-\frac{1}{\rho} \nabla P + {\bf g},
\label{eq:momcon}
\end{equation}
where $P$ is the gas pressure and ${\bf g}$ is the net
external force on a parcel of gas, here assumed to be only
due to gravity.
The equation of total energy conservation,
\begin{equation}
\frac{\partial}{\partial t} \big(\frac{\rho u^2}{2} + \frac{3P}{2} \big) +
\nabla \cdot \Big[ {\bf F}_{H} + {\bf F}_{C} + \rho {\bf u}
\big(\frac{u^2}{2} + \frac{5P}{2\rho} - \frac{GM_{\odot}}{r}
\big) \Big] = -\rho^{2} \Phi(T),
\label{eq:encon}
\end{equation}
contains several terms that arise from the ionized nature of the
near-Sun plasma. The solar corona exhibits a temperature of about 
$10^6$\,K, which is at least two orders of magnitude higher than the
temperature at the base of the atmosphere (that is, the photosphere
and chromosphere). A major unsolved problem of solar physics is to
explain
what physical processes lead to such large amounts of
energy into the coronal plasma.
Significant progress has been made, though, by constraining
the input energy flux density ${\bf F}_H$ empirically, even
though the exact physical origin of this heating is not known.

Some of the energy deposited into the corona is transported
downward to the lower atmosphere via heat conduction
(that is, the radial component of ${\bf F}_{C}$ is negative),
and some of it is converted back and forth between
kinetic energy, thermal energy, and gravitational
potential energy. (See the final three terms in square brackets in
Eq.~(\ref{eq:encon}), where $G$ is the gravitational constant and
$M_{\odot}$ is the mass of the Sun.)
Some of this energy also is lost in the form of radiation,
because many of the free electrons that exist in an ionized
plasma undergo collisions with bound atoms and liberate a
fraction of their energy to photon emission.
The radiative loss function $\Phi(T)$ encapsulates
the elemental composition and atomic physics of the radiating
coronal plasma as a function of temperature $T$.

Equations~(\ref{eq:mascon})--(\ref{eq:encon}) can be simplified in several
ways without sacrificing realism. We can neglect the partial time derivatives
and restrict our analysis to time-steady solutions of a continuously
expanding solar wind. Also, the vector terms can be expressed in spherical
polar coordinates, assuming that the variations exist only along the radial
direction
$r$ and that all vectors have nonzero components only in this
direction. For example, the radial component of the gravitational
acceleration ${\bf g}$ is simply $g_{r} = -GM_{\odot}/r^{2}$.

Very near the Sun, the assumption of radial flow is not
accurate because of the complex multipole structure of the
solar magnetic field (see Fig.~1).
At larger distances, though, the radial flow of the solar
wind has been largely confirmed by spacecraft
measurements.\cite{Hu72,Jk97}
However, the Sun's slow rotation (once every 27 days) causes
the flow direction and the magnetic field direction to become
misaligned in interplanetary space.
Because of the high conductivity of the solar wind plasma,
the magnetic field lines become ``frozen in'' to the flow, that is,
the magnetic field becomes a passive tracer of the flow, like drops
of ink in a flow of water. A possible outdoor demonstration of this
effect can be performed with a persistent, rotating source of water flow,
like a lawn sprinkler. Despite the fact that all of the water droplets are
flowing radially away from the center, a snapshot at any time
shows them arranged in a spiral ``streakline.'' This field is completely
analogous to the Parker spiral magnetic field pattern in the solar wind,
which carries the imprint of the Sun's rotation, but still channels the
particle flow to be radial.

For the useful assumption of radial flow, the mass
conservation equation is expressed in spherical symmetry as
\begin{equation}
\frac{1}{r^2} \frac{d}{dr} (\rho u r^{2}) = 0.
\label{eq:masconR}
\end{equation}
Because the radial derivative of $\rho u r^2$ is zero, this
quantity is constant. 
We thus define the total mass loss rate from the entire Sun
(in units of kg~s$^{-1}$) as $\dot{M} \equiv 4 \pi \rho u r^{2}$,
where the factor of $4\pi$ comes from integrating over the full
solid angle of the spherical Sun.
The energy conservation equation is written as
\begin{equation}
\frac{1}{r^2} \frac{d}{dr} \Big\{ r^{2}
\big[ F_{H} + F_{C} + \rho u
\big(\frac{u^2}{2} + \frac{5P}{2\rho} - \frac{GM_{\odot}}{r}
\big) \big] \Big\} = -\rho^{2} \Phi(T) ,
\label{eq:enconr2}
\end{equation}
with the radial components of vectors written as
scalars with the same notation.

The gas pressure $P$ can be eliminated from the momentum
equation by applying the ideal gas law, assuming the fluid
consists of a single particle species with mass $m$,
\begin{equation}
P = \frac{\rho k T}{m} \equiv \rho a^{2},
\end{equation}
where $k$ is Boltzmann's constant
and $a$ is an effective sound speed.
For a hydrogen plasma, $m$ is essentially the proton mass.
An added assumption that simplifies the subsequent analysis
is that the hot corona is isothermal, that is, that after
the coronal heating takes hold, the $\sim$10$^6$\,K temperature
remains roughly constant as a function of radius. It is known from
spacecraft measurements that the plasma temperature drops only by a
factor of 10 from the inner corona ($r \approx 1.5 \, R_{\odot}$)
to the orbit of the Earth ($r \approx 215 R_{\odot} = 1$\,AU),
where $R_{\odot}$ is the solar radius.\cite{Hu72}
Therefore, for the acceleration region of the solar wind (1.5 to
$10 R_{\odot}$), the isothermal approximation seems sufficiently
valid.

If we substitute these conditions into
Eq.~(\ref{eq:momcon}) and use the mass conservation equation, we
obtain
\begin{equation}
\big(u - \frac{a^2}{u} \big) \frac{du}{dr} = 
\frac{2a^2}{r} - \frac{GM_{\odot}}{r^2}.
\label{eq:transs}
\end{equation}
It is noteworthy that the momentum conservation equation is
now a true equation of motion because the mass density
$\rho$ no longer appears.

\section{The Parker Solar Wind Problem}\label{sec4}

The fluid in a steady-state stellar wind accelerates
from rest to an asymptotic ``coasting'' speed far from the
star, where the star's gravity has become negligible.
This situation can only be maintained by a gradual
transition from a hydrostatic force balance close to
the star (that is, where inward and outward forces cancel)
to a net outward force at larger distances.\cite{Vinf}
Parker\cite{P58} recognized that this transition occurs
naturally for a hot (million K) corona, where the gradient of the
large gas pressure plays the role of the increasing outward force.
Equation~(\ref{eq:transs}) shows the primary manifestation of
the gas pressure gradient as the first term on the
right-hand side, which for a constant $a$ eventually
must overtake the more steeply decreasing gravity term
and result in a net positive (outward) acceleration.
Interestingly, the dynamics described by Eq.~(\ref{eq:transs})
does not depend on how the corona is heated, but
merely on the fact that it is heated.

Parker also noticed that Eq.~(\ref{eq:transs}) exhibits a
potential singularity at the ``sonic point,'' $u=a$,
because when this condition applies, the term in parentheses
on the left-hand side is zero, and for an arbitrary radius
(that is, a finite value for the right-hand side) the first
derivative of the velocity $du/dr$ must be infinite.
However, there is one specific value for $r$ where the
right-hand side is zero as well.
If the sonic point occurs at the critical radius
$r_{c} = GM_{\odot} / (2 a^{2})$, then $du/dr$ may remain
finite and the wind solution remains physically realistic.
Mathematically, this solution represents an X-type singular point,
at which two solution trajectories in $(r,u)$ space intersect
with slopes of opposite signs and other solutions are
hyperbolic about this point.
The joint set of conditions $r = r_c$ and $u = a$
often is called the Parker critical point.

There are two possible solutions that pass through the
critical point: one representing a continuously
accelerating outward flow of gas (the wind), and one
representing an outwardly decelerating, but inward
flow of gas (steady spherical accretion).
Parker's wind solution was criticized initially for being
too ``finely tuned'' because it seemed unlikely that a wind
would naturally want to accelerate through the sonic point
exactly at $r = r_c$.
However, it has been noticed recently that Parker's critical
solution is the only truly stable wind solution to
Eq.~(\ref{eq:transs}), and all other outwardly flowing
solutions are unstable.\cite{Ve01}
Note that six years before Parker, Bondi\cite{Bo52}
recognized that the inward accretion solution also could 
represent real astrophysical flows.
It was also found that this solution, like Parker's, is
a stable attractor and represents the maximum amount
of mass that can be consumed (in steady state) from an external
source.\cite{Ch90}

Equation~(\ref{eq:transs}) is a first-order ordinary differential
equation that is separable.
The integration of the left side from $a$ to an arbitrary $u$ and
the right side from $r_c$ to an arbitrary $r$ yields an implicit
transcendental equation for $u$ and $r$:
\begin{equation}
\label{this}
(u^{2} - a^{2}) - a^{2} \ln \big(\frac{u^2}{a^2} \big) = 
4a^{2} \ln \big(\frac{r}{r_c} \big) + 2 GM_{\odot}
\Big(\frac{1}{r} - \frac{1}{r_c} \Big).
\end{equation}
We rearrange terms and define the dimensionless variable
$y \equiv (u/a)^2$, so that Eq.~(\ref{this}) becomes
\begin{equation}
\ln y - y = \ln D(r),
\end{equation}
where
\begin{equation}
D(r) = \big(\frac{r}{r_c} \big)^{-4} 
\exp \big[ 4 \big(1 - \frac{r_c}{r} \big) - 1 \big].
\end{equation}
Thus, using Eqs.~(\ref{eq:briggs1}) and (\ref{eq:briggs2}),
the Parker/Bondi solutions have the general analytic solution
$y = -W[-D(r)]$.
For all values of $r$, $D(r)$ ranges between 0 and $1/e$,
so we must choose between the two branches $W_0$ and
$W_{-1}$ (see Fig.~2).
Additionally, the full solution involves one choice below
the critical point and the opposite choice above it.
The accelerating Parker solar wind solution is given
specifically by
\begin{equation}
u^{2} = 
\begin{cases}
-a^{2} W_{0}[-D(r)] &
\text{$r \leq r_{c}$} \\
-a^{2} W_{-1}[-D(r)] &
\text{$r \geq r_{c}$}
\end{cases}
\label{eq:windsol}
\end{equation}
and the opposite choices must be made to obtain the
Bondi accretion solution.

Figure~3 shows a set of solutions for the Parker solar wind
with six choices for the constant sound speed $a$.
These solutions are compared to curves showing empirical
(that is, observationally derived) speeds for the fastest and
slowest types of solar wind flow that have been seen.
Because our only direct measurements of the wind speed have
been exterior to the orbit of Mercury ($r > 60 R_{\odot}$),
indirect methods are needed to determine the wind speed at
distances closer to the Sun.
The analysis of ultraviolet photons emitted from the corona
has provided new ways of probing the solar wind's
acceleration,\cite{Cr02} but a more traditional method
is to measure the density of particles in the corona
and use mass conservation, that is, the steady state version of
Eq.~(\ref{eq:masconR}), to compute the wind speed.
The density can be measured by observing the linear polarization
of Thomson-scattered visible light in the corona; the degree
of polarization is directly proportional to the number of
free electrons along the line of sight.

In the following we give a simple parameterization\cite{SG99} of the
radial dependence of density as observed in the source regions
of the fast and slow components of the solar wind (at the
minimum of the Sun's 11-year magnetic cycle):
\begin{subequations}
\label{above}
\begin{eqnarray}
\rho \, \mbox{(fast)} &\approx& 2.37 \times 10^{-19} 
\big(\frac{1}{x^2} + \frac{5.9}{x^3} +
\frac{650}{x^9} \big)\,\mbox{g}/\mbox{cm}^{3} \\
\rho \, \mbox{(slow)} &\approx& 6.21 \times 10^{-19} 
\big(\frac{1}{x^2} + \frac{13}{x^3} +
\frac{480}{x^6} \big)\,\mbox{g}/\mbox{cm}^{3},
\end{eqnarray}
\end{subequations}
where $x = r/R_{\odot}$.
The wind speed at any radius is thus proportional to
$\rho^{-1} r^{-2}$ times a normalization constant that is given
by specifying the measured wind speed at 1\,AU.
Note that at large distances the dominant terms in Eq.~(\ref{above})
will be the $1/x^{2}$ terms, and thus $u \propto \rho^{-1} r^{-2}$
at large distances will approach a constant coasting speed.

The density is generally higher in the slower component
of the solar wind, which emerges mainly from bright ``streamers''
around the solar equator and reaches speeds at 1\,AU of
300 to 500\,km/s.
The density is lowest in the fast solar wind that emerges
mainly from dark ``coronal holes'' at the north and south
poles and reaches speeds at 1\,AU of 600 to 800\,km/s.\cite{Ma01}
Figure~3 shows that the acceleration of the slow wind has a
very similar shape to the analytic solutions given in
Eq.~(\ref{eq:windsol}).
The fast wind has a slightly steeper profile in
the corona because this plasma is not isothermal, and because it
also flows slightly ``super-radially'' (that is, the magnetic field
over the poles flares out like a trumpet and the equations are not
represented exactly by spherical symmetry; see Fig.~1). However,
much of the essential physics of solar wind acceleration remains
encapsulated in the radial, isothermal problem.

One practical benefit of having an analytic expression for
$u(r)$ is being able to easily find asymptotic expansions
for various limiting cases.
In the nearby vicinity of the critical point, that is, for
$|1 - (r_{c}/r)| \ll 1$, the series expansions given by
Eqs.~(\ref{eq:W0approx}) and (\ref{eq:W1approx}) can be
used in conjunction with the series expansion of $D(r)$ about
the critical point to obtain a single expression for
radii near $r_c$:
\begin{equation}
u \approx a \, \sqrt{3 - \frac{2r_c}{r}},
\qquad
\mbox{when}\ r \approx r_{c}.
\end{equation}
Other expansions can be used to obtain approximations
for $r \ll r_{c}$ and $r \gg r_{c}$.

\section{The Mass Flux Problem}

The above solution (Eq.~[\ref{eq:windsol}]) for the
solar wind speed
$u(r)$ is only half of the problem.
Because the mass density $\rho$ was eliminated from the
equation of motion, we know how fast the gas is accelerating,
but we do not know how much gas is being ejected.
The determination of the solar wind mass loss rate $\dot{M}$
is the second half of the problem which, interestingly, also is
addressable using the Lambert $W$ function.

The Sun is observed to lose mass at a rate of approximately
$10^{-14}$ solar masses per year ($M_{\odot}$/yr).
This unconventional unit is
useful because it can be compared easily to a firm upper
limit derivable by dividing the mass of the star by its
lifetime.
For the Sun, with an expected main-sequence lifetime of
about $10^{10}$ years, this upper limit is of order
$10^{-10}\,M_{\odot}$/yr.
Thus, the solar wind is expected to drain away no more
than one ten-thousandth of the Sun's mass over the next
few billion years.
(Some hotter stars lose mass at much higher rates, with
the wind having a substantial impact on the star's late
stages of evolution.\cite{LC99})

There is still not universal agreement about what determines
the Sun's mass loss rate.\cite{Ha82,Wi88,LM99,Fi03}
Our analytic solutions apply to only one of the
several suggested mechanisms.
In this class of radiative energy balance models,
first outlined in detail by Hammer,\cite{Ha82}
$\dot{M}$ is determined at the base of the corona by the
interplay between the heating and cooling terms in
Eq.~(\ref{eq:enconr2}), the equation of energy conservation.
Because we have solved for $u(r)$, the determination
of the mass loss rate requires only the solution for the
density at a single radius.

To simplify Eq.~(\ref{eq:enconr2})
further, the solar atmosphere can be considered to consist
of two concentric layers: the cool ($\sim$10$^4$\,K),
high density chromosphere, and the overlying hot
($\sim$10$^6$\,K), low density corona.
The transition between these layers has been observed to
be exceedingly thin---about 0.1\% of a solar radius---so that
the radial derivative in Eq.~(\ref{eq:enconr2}) can be
expressed as a simple difference of quantities above
and below the transition zone.
Because of the relative thinness of this zone,
we can ignore both the small change in the gravitational
potential energy between the two layers and the
spherical divergence, that is, the $r^2$ terms inside and
outside the braces.
Also, the kinetic energy term $u^{2}/2$ can be ignored
because the solar wind speed has been seen to be negligibly
small (that is, very subsonic) at the solar surface.
Finally, the coronal heating term itself, $F_H$, is
ignorable because we are concerned with layers below
where the majority of the heat is deposited.
Thus there are only three dominant terms in the energy
balance:
\begin{equation}
\frac{d}{dr} (F_{C} + 5nukT) = 
-\rho^{2} \Phi(T),
\end{equation}
where for convenience we rewrite the mass density $\rho$ as the
product of the particle mass $m$ and a number density $n$, that is, number of
particles per unit volume. In summary, at the coronal base the heat
is conducted downward from where it is initially deposited, some of
it resides at the base as enthalpy, and the remainder is lost as
radiation.

The steady state balance of mass and momentum across the
thin transition zone also demands that the products
$nu$ (mass flux) and $nT$ (gas pressure) remain roughly
constant.
This mass flux constraint is used, together with an
empirical form\cite{RTV,Wa94} for the radiative loss
function $\rho^{2} \Phi \equiv n^{2} A T^{-1/2}$
(where $A = 1.9 \times 10^{-32}$\,W\,m$^3$\,K$^{1/2}$),
to obtain
\begin{equation}
\label{this2}
\frac{dF_C}{dr} + 5nuk \frac{dT}{dr} = 
-A n^{2} T^{-1/2}.
\end{equation}
The differential equation (\ref{this2}) is transformed by
multiplying both sides by the heat conductive flux $F_C$.
Note, though, that it is advantageous to multiply the
left-hand side by $F_C$ itself and to multiply the
right-hand side by the definition of the classical
conductive flux,
\begin{equation}
F_{C} \equiv -\kappa T^{5/2} \frac{dT}{dr},
\label{eq:spitharm}
\end{equation}
where $\kappa$ is the Spitzer-H\"{a}rm\cite{Sp62} heat
conductivity in an ionized plasma, which has a
value of $8.8 \times 10^{-12}$\,W\,m$^{-1}$\,K$^{-7/2}$
for the range of densities and temperatures of the corona.
We rearrange and divide all terms by a factor of
$dT/dr$ and obtain the following form of the energy balance
equation:
\begin{equation}
\xi F_{C} + F_{C} \frac{dF_C}{dT} = \psi,
\label{eq:xipsi}
\end{equation}
where the quantities $\xi = 5nuk$ and
$\psi = n^{2} \kappa A T^2$ are assumed to be constant across
the thin transition zone.

The above form (Eq.~[\ref{eq:xipsi}]) of the energy equation is
separable and integrable with $T$ and $F_C$ as the independent and
dependent variables, respectively.
Once integrated across the transition zone, though, the
full equation contains terms for $T$ and $F_C$ in both the
upper and lower layers.
The terms corresponding to the lower (chromospheric)
layer can be neglected because the values of both $T$ and
$F_C$ are several orders of magnitude smaller in
comparison to their counterparts in the upper (coronal) layer.
Thus, the integrated transcendental equation relates the
values of $T$ and $F_C$ at the coronal base to one another
and is independent of their values in the chromosphere:
\begin{equation}
F_{C} + \xi T + \frac{\psi}{\xi} \ln \big(1 -
\frac{\xi F_C}{\psi} \big) = 0.
\end{equation}
We note that both $\xi$ and $\psi$ contain the number
density at the coronal base (which we call $n_0$). Hence,
the Lambert $W$ function can be used to solve for
this quantity, with
\begin{equation}
\label{n0}
n_{0} = \frac{5 u_{0} k F_C}
{\kappa A T^2 [ 1 + W(\omega) ]}
\end{equation}
and the argument $\omega$ of the $W$ function is
\begin{equation}
\omega = - \exp \big(-\frac{25 u_{0}^{2} k^{2}}{\kappa A T}\big) -
1.
\end{equation}

In this formulation, the only ``free'' variables are
$u_0$, $F_C$, and $T$ (all evaluated at the coronal base).
Just as in the Parker solar wind application, the argument
$\omega$ falls between $-1/e$ and 0, thus making
the choice between the $W_0$ and $W_{-1}$ branches necessary.
In this case, though, the physical choice ($W_{-1}$) is
apparent because the $W_0$ branch gives a negative density.

To use the above solution (Eq.~[\ref{n0}]) to compute the
mass loss rate, we use the results of Sec.~\ref{sec4} to fix
$u_0$ for a given isothermal Parker wind model. A realistic median
value of $T$ is $1.2 \times 10^6$\,K, which has an outflow
speed $u_0$ of 0.96\,km/s at $r = R_{\odot}$.
To estimate the applicable values of $F_C$ at the
coronal base, Eq.~(\ref{eq:spitharm}) can be solved assuming
a finite-difference temperature gradient across the thin
transition zone.
A thickness $\Delta r$ of 0.001~$R_{\odot}$ gives
values of $F_C$ of order --1000 W\,m$^{-2}$.
In more accurate models,\cite{Wa94} though, the temperature
gradient is a bit less steep at the top of the transition
zone, and $F_C$ ranges between --50 and
--200\,W\,m$^{-2}$.
For this range, Eq.~(\ref{n0}) for $n_0$ yields
values of the number density between
$6 \times 10^{13}$ and $3 \times 10^{14}$\, m$^{-3}$.
The mass flux, integrated around the whole sphere, is then
$\dot{M} \equiv 4 \pi m_{p} n_{0} u_{0} R_{\odot}^{2}$,
and the resulting values range between
$9 \times 10^{-15}$\,$M_{\odot}$/yr and
$4 \times 10^{-14}$\,$M_{\odot}$/yr.
The observed solar mass loss rate is observed to vary
between about $2 \times 10^{-14}$\,$M_{\odot}$/yr at the
solar minimum and a few times that at the solar maximum.
If we take into consideration the large number of approximations
we have applied, we conclude that the agreement is good.

\section{Conclusions}

We have presented new analytic solutions
to two simple problems in solar wind physics.
The Lambert $W$ function used in these solutions was defined
and publicized only about a decade ago, but it has rapidly
become a convenient tool for mathematical physicists.
The elegance of explicit solutions to equations thought
previously to be expressible only implicitly is clear, but
there also are many practical benefits to having explicit
solutions as well.

There are other potential applications of the Lambert
$W$ function in solar and space physics.
A transcendental equation solvable in terms of $W$ arises in a
calculation of the electric potential drop that exists
between the Sun and the edge of the solar system.\cite{MV99}
(An ionized plasma exhibits local charge neutrality because
of electrostatic screening, but for solar wind particles in the Sun's
gravitational field this neutrality is possible only by setting up a radially
varying electric field.) Functions with temperatures $T$ appearing both
inside and outside exponents occur
when calculating the energy distributions of solar photons (for example, the
Planck blackbody function) and electrons in excited atoms (for example, the
Saha ionization equation). The Lambert $W$ function can thus be used in a
variety of ways when the need to solve for $T$ arises. Further applications
are expected to clarify the physics of many types of systems.

\begin{acknowledgments}
Valuable and encouraging discussions with Adriaan van Ballegooijen
and Willie Soon substantially improved this paper. This work is
supported by the National Aeronautics and Space Administration
under grants NAG5-11913 and NAG5-12865 to the Smithsonian
Astrophysical Observatory.
\end{acknowledgments}

\newpage

\section*{Figure Captions}

\begin{figure}[ht]
\begin{center}
\includegraphics{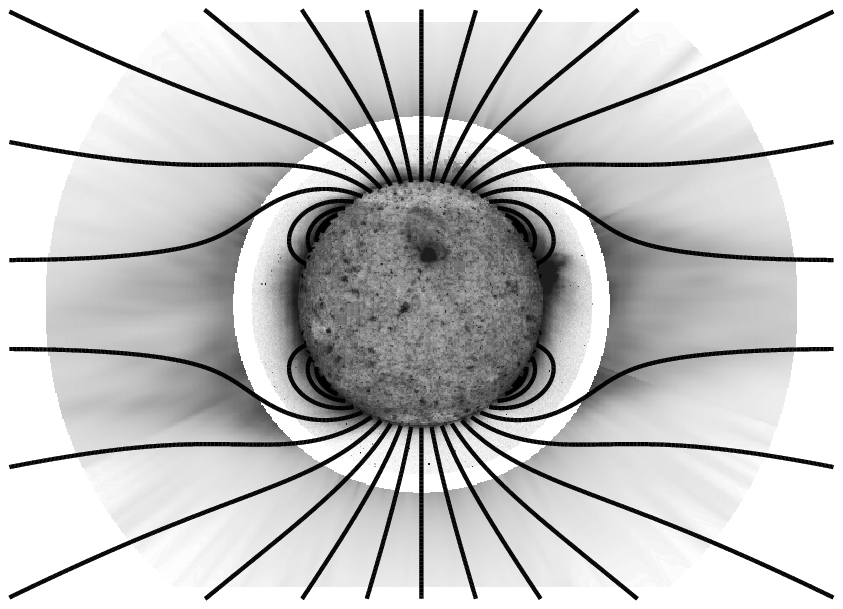}
\caption{The solar corona on August 17, 1996, with bright
regions plotted as dark.
The inner image was taken by the EIT
(Extreme-ultraviolet Imaging Telescope)
instrument on Solar and Heliospheric Observatory (SOHO), and is 
sensitive to the ultraviolet
emission of Fe$^{+11}$ ions at temperatures of about $10^6$\,K.
The outer image was taken by the UVCS
(Ultraviolet Coronagraph Spectrometer)
instrument on SOHO by blocking out the bright disk to see the
much dimmer ultraviolet emission of O$^{+5}$ ions at temperatures
exceeding $10^8$\,K. The magnetic field lines are from a model
of the corona at the minimum of its 11-year activity
cycle.\cite{Ba98}}
\end{center}
\end{figure}

\begin{figure}[ht]
\begin{center}
\includegraphics{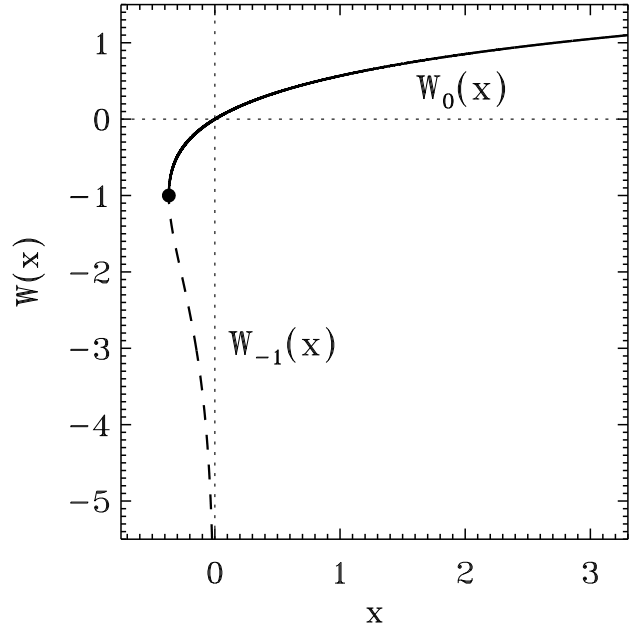}
\caption{The two real branches of the Lambert $W$ function.}
\end{center}
\end{figure}

\begin{figure}[ht]
\begin{center}
\includegraphics{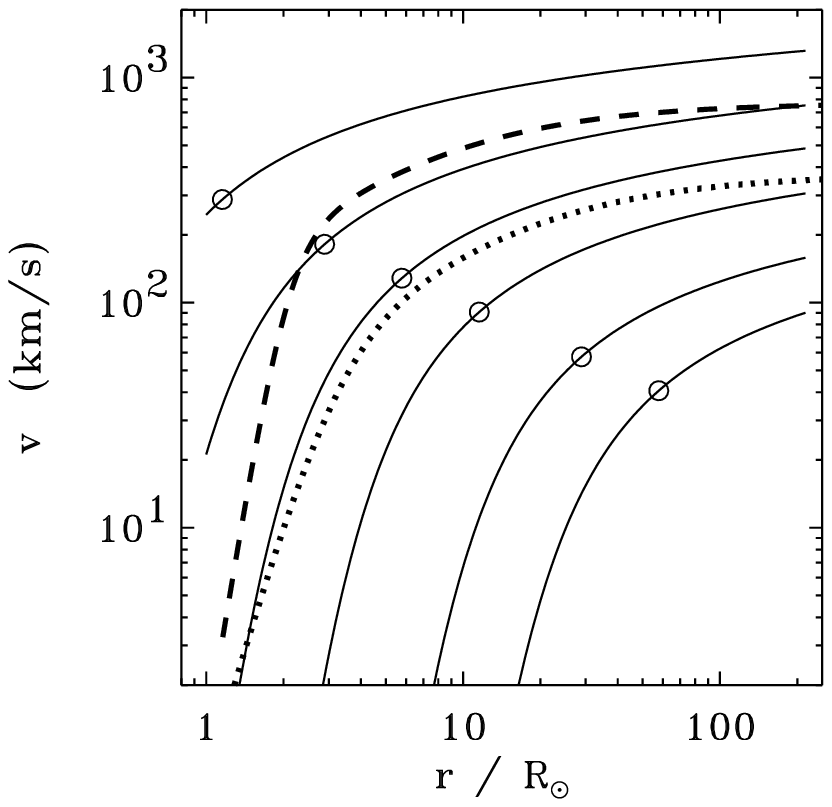}
\caption{Analytic solutions of the isothermal Parker
solar wind equation, plotted as outflow speeds versus
heliocentric distance (in units of a solar radius;
$R_{\odot} = 6.96 \times 10^{8}$\,m).
Individual solutions (thin solid lines) are
labeled with the locations of the critical point (circles).
From bottom to top, the modeled coronal temperatures are
0.1, 0.2, 0.5, 1, 2, 5\,MK, respectively. Shown for comparison
are observationally constrained wind speeds for polar
(thick dashed line) and equatorial (thick dotted line) flow
at solar minimum.}
\end{center}
\end{figure}


\begin{thebibliography}{99}
\bibitem{Vinf}
After the wind is accelerated past the star's escape velocity,
the magnitude of the net outward force need not be large.
Indeed, in the solar case both the inward force of gravity
and the outward driving force tend to zero as
$r \rightarrow \infty$, and thus the wind approaches
an asymptotic coasting speed or terminal velocity.
\end{thebibliography}
\end{document}